\documentclass[letter]{aa} % for the letters

\usepackage[utf8]{inputenc}
\usepackage{graphicx}
\usepackage{txfonts}
\graphicspath{{Figures/}}
\usepackage{ulem}

\usepackage{natbib,twoopt}
\usepackage[breaklinks=true]{hyperref} %% to avoid \citeads line fills
\bibpunct{(}{)}{;}{a}{}{,}             %% natbib format for A&A and ApJ
\makeatletter
  \newcommandtwoopt{\citeads}[3][][]{\href{http://adsabs.harvard.edu/abs/#3}%
    {\def\hyper@linkstart##1##2{}%
     \let\hyper@linkend\@empty\citealp[#1][#2]{#3}}}
  \newcommandtwoopt{\citepads}[3][][]{\href{http://adsabs.harvard.edu/abs/#3}%
    {\def\hyper@linkstart##1##2{}%
     \let\hyper@linkend\@empty\citep[#1][#2]{#3}}}
  \newcommandtwoopt{\citetads}[3][][]{\href{http://adsabs.harvard.edu/abs/#3}%
    {\def\hyper@linkstart##1##2{}%
     \let\hyper@linkend\@empty\citet[#1][#2]{#3}}}
  \newcommandtwoopt{\citeyearads}[3][][]%
    {\href{http://adsabs.harvard.edu/abs/#3}
    {\def\hyper@linkstart##1##2{}%
     \let\hyper@linkend\@empty\citeyear[#1][#2]{#3}}}
\makeatother
\newcommand{\St}{\ensuremath\mbox{St}}
\newcommand{\Ro}{\ensuremath\mbox{Ro}}

\begin{document}

  \title{Reduction of dust radial drift by turbulence in protoplanetary disks}

   \author{
        Fabiola Antonietta Gerosa\inst{1},
        J\'{e}r\'{e}mie Bec
          \inst{2,3},
          H\'{e}lo\"{\i}se M\'{e}heut\inst{1}
          \and Anand Utsav Kapoor\inst{1}
          }
    \authorrunning{F.A.\ Gerosa et al.}
   \institute{Université Côte d’Azur, Observatoire de la Côte d’Azur, CNRS, Laboratoire Lagrange, Nice, France.\\
              \email{fabiola.gerosa@oca.eu}
         \and
             Universit\'{e} C\^{o}te d'Azur, CNRS, Institut de Physique de Nice, France.
         \and
             Université Côte d'Azur, Inria, CNRS, Calisto team, Sophia Antipolis, France.
             }

   \date{}

% \abstract{}{}{}{}{}
% 5 {} token are mandatory
 
  \abstract
   {Dust particles in protoplanetary disks rotate at velocities exceeding those of the surrounding gas due to a lack of pressure support. Consequently, they experience a head-wind from the gas 
   that drives them toward the central star. Radial drift occurs on timescales much shorter than those inferred from disk observations or those required for dust to aggregate and form planets. Additionally, turbulence is often assumed to amplify the radial drift of dust in planet-forming disks when modeled through an effective viscous transport.
   However, the local interactions between turbulent eddies and particles are known to be significantly more intricate than in a viscous fluid.}
   {Our objective is to elucidate and characterize the dynamic effects of Keplerian turbulence on the mean radial and azimuthal velocities of dust particles.}
   {We employ 2D shearing-box incompressible simulations of the gas, which is maintained in a developed turbulent state while rotating at a sub-Keplerian speed. Dust is modeled as Lagrangian particles set at a Keplerian velocity, therefore experiencing a radial force toward the star through drag.}
   {Turbulent eddies are found to reduce the radial drift, while simultaneously enhancing the azimuthal velocities of small particles. This dynamic behavior arises from 
   the modification of dust trajectories due to turbulent eddies.
   }
   {}

   \keywords{Dust drift -- Turbulence -- Planet formation -- Planetesimals
               }

   \maketitle
%
%________________________________________________________________

\section{Introduction}

Radial drift stands out as a pivotal process in the evolution of dust within planet-forming disks, playing a crucial role in planet formation. Dust drift arises from the sub-Keplerian rotation of the gas, exerting drag forces  slowing down the inherently Keplerian motion of dust. The resulting depletion of dust angular momentum triggers its radial displacement toward the central star on a timescale significantly shorter than the disk's lifetime. 
This process is seemingly in contradiction with the observed presence of dust in much older disks. Moreover, planet formation is expected to occur over timescales on the order of a million years and needs a large amount of dust left in the disk to form the observed massive objects. However, this timescale is significantly longer than the radial drift timescale, leading to an inconsistency in planet formation theories, known as the radial drift barrier \citep{weidenschilling1977,nakagawa1986}. Consequently, planetesimal formation is believed
to be localized in specific regions where radial drift is arrested,
and dust concentrates. \\
Recent observational evidences suggest that disk turbulence exhibits lower amplitude than previously believed \citep{flaherty2015,flaherty2018,villenave2022}. This is consistent
 with the saturated non-linear state obtained 
from various hydrodynamical instabilities, such as the vertical shear instability \citep{arlt2004,nelson2013} or the convective overstability \citep{klahr2014,lyra2014}, rather than situations displaying  strong turbulence \citep[e.g., through the magnetorotational instability,][]{balbus1991}. A weak turbulence modifies our understanding of turbulent particle concentration. In particular, rotation and shear become non-negligible compared to the strength of turbulence, resulting in a reshape of the turbulent eddies. In this context, turbulence cannot be treated as homogeneous and isotropic. Our previous work \citep{gerosa2023} demonstrates that rotation, through the Coriolis force, can counterbalance the ejection of particles from turbulent eddies. This phenomenon fosters clustering pf dust and potential planetesimal formation within anticyclonic eddies, in contrast with earlier studies that focused on
lower rotation rates \citep{cuzzi2001,pan2011}. Furthermore, it shows that Keplerian turbulence cannot be simplistically approximated as a purely diffusive process acting on particles.\\
Turbulence is frequently invoked for its ability to transport gas angular momentum, often modeled as a viscosity \citep{shakura1973,lesur2021}. This paradigm leads to gas radial motion, consequently amplifying dust radial drift through drag in an accreting disk \citep{takeuchi2002} . However, such a mean-field approach neglects the local modifications that turbulent eddies can impose on particle trajectories. The nuanced understanding of non-isotropic turbulence and its interplay with dust dynamics, as elucidated in \citet{gerosa2023}, prompts an investigation into whether Keplerian turbulence additionally alters dust drift. Turbulent gas could potentially slow down or even halt the radial drift of dust, thus overcoming the need for an azimuthally extended pressure bump, which is the conventional solution to this barrier \citep{whipple1972, pinilla2012}.
\\
The aim of this study is to investigate the impact of gas turbulence on the radial and azimuthal drift of dust. We employ 2D numerical simulations of forced Keplerian turbulence within a shearing box and track the dynamics of drifting Lagrangian particles in this flow, characterized by the absence of mean radial velocity. Our findings reveal how turbulence plays a crucial role in diminishing particles radial drift, while concurrently enhancing their azimuthal drift.

\section{Model and numerical methods}\label{sec:numerics}

\subsection{Gas}
We assume that the gas is incompressible; thus, its dynamics are governed by a divergence-free velocity field $\vec{v}$ that satisfies the Navier--Stokes equation, with the inclusion of the external gravitational potential from the star. Additionally, we consider the gas flow to be two-dimensional. We consider a small region located at distance $r_0$ from the star, rotating at a rate $\Omega$ along with the gas, and we employ the periodic shearing box approximation. If $(r,\theta)$ are the polar coordinates centered onto the star, we define the local coordinates $x=r-r_0$ and $y=r_0\theta$. The total gas velocity solves:
\begin{equation}
 \partial_t \vec{v} + \vec{v}\cdot\vec{\nabla} \vec{v} = - \frac{1}{\rho_{\rm g}}\vec{\nabla} p + \nu \vec{\nabla}^2 \vec{v} -2\vec{\Omega}\times \vec{v} + 3\Omega^2\, x\, \vec{e}_x + \vec{f} -\vec{D}, 
 \label{eq:NS}
\end{equation}
where $\rho_{\rm g}$ and $\nu$ represent the gas mass density and kinematic viscosity, respectively. The computational domain has size $L_x\times L_y$, with an aspect ratio of $L_y/L_x=4$ to limit spurious geometrical effects at large $\Omega$. The fluctuating velocity of the gas $\vec{u} = \vec{v} +\tfrac{3}{2}\Omega\,x\,\vec{e}_y$ is assumed periodic in the frame distorted by the mean shear. We conduct numerical simulations with $256\times1024$ collocation points using the open-source pseudo-spectral solver \textsc{Snoopy} \citep[see][]{lesur2005}.

The turbulent flow is maintained in a statistically steady state by introducing both a forcing term, $\vec{f}$, and a large-scale dissipation $\vec{D} = \gamma\,(\vec{v} +\tfrac{3}{2}\Omega\,x\,\vec{e}_y)$. The prescribed force is random, Gaussian, homogeneous, isotropic in the sheared frame, with a zero mean, white noise in time, and with spatial correlations concentrated at large scales. This implementation allows us to investigate a generalized form of 2D turbulence. Although the resulting flow exhibits some similarities with gas subjected to the subcritical baroclinic instability \citep{lesur2010}, investigating the transition to a turbulent state through instabilities is beyond the scope of this paper. The forcing has a fixed amplitude, so that the mean injection rates of kinetic energy $\varepsilon_{\rm I}$ and enstrophy $\eta_{\rm I}$ are prescribed. This specifies a forcing length-scale $\ell_{\rm f} = (\varepsilon_{\rm I}/\eta_{\rm I})^{1/2}$ and timescale $\tau_{\rm f}=(\eta_{\rm I})^{-1/3}$, which remain constant across all runs. We can also define the small eddy turnover time $\tau_\omega=\langle\omega^2\rangle^{-1/2}$, where $\omega=\partial_x u_y-\partial_y u_x $ is the vorticity of the gas turbulent fluctuations. Using this timescale, we non-dimensionalize the rotation rate $\Omega$, defining the Rossby number $\Ro=1/(\tau_{\omega}\Omega)$. It serves as a measure of turbulence strength relative to rotation. For our study, we have selected values of $\Ro$ in the range of $[0.1,10]$ (see Appendix~\ref{app:run} for details on the simulation parameters). 
Notice that at small $\Ro$, rotation becomes predominant over turbulence. The effects of such weak turbulence on particle dynamics have been understudied until now. 
 
\subsection{Dust}

The position and velocity $(\vec{R}_{\rm p},\vec{U}_{\rm p})$ of a particle in the rotating and sheared frame solve
\begin{eqnarray}
    \frac{d\vec{R}_{\rm p}}{dt}=&&\hspace{-17pt}\vec{U}_{\rm p}+\frac{3}{2}\Omega R_{{{{\rm p},x}}}\,\vec{e}_y,  \\
    \frac{d\vec{U}_{\rm p}}{dt}=&&\hspace{-17pt}-\frac{1}{\tau_{\rm p}}\left[\vec{U}_{\rm p}-\vec{u}(\vec{R}_{\rm p},t)\right]-2\vec{\Omega}\times\vec{U}_{\rm p}+\frac{3}{2}\,\Omega\,U_{{{\rm p},x}}\,\vec{e}_y-\epsilon\, \vec{e}_x.
    \label{eq:particles}
\end{eqnarray}
The equation governing the motion of dust particles is derived in Appendix \ref{app:shearingbox}.
The first term on the right-hand side of Eq.~(\ref{eq:particles}) comes from the drag with the gas. It involves the particle stopping time $\tau_{\rm p}$ defined 
in Epstein regime as $\tau_{\rm p}=(8/\pi)^{1/2}\rho_{\rm p} a/(\rho_{\rm g} c_{\rm s})$, with $\rho_{\rm p}$ denoting the particle material mass density, $a$ its radius and $c_{\rm s}$ the speed of sound. We non-dimensionalize $\tau_{\rm p}$ by defining the Stokes number $\St=\tau_{\rm p}\,\Omega$.
The subsequent term appearing in Eq.~(\ref{eq:particles}) accounts for the Coriolis force. The last term represents the adjustment for the difference in azimuthal velocity between gas and dust in the shearing box framework. It involves the parameter $\epsilon=(\Omega_{\rm K}^2-\Omega^2)r_0$, where $\Omega_{\rm K}$ denotes the particle Keplerian rotation rate. When scaled relative to the two other fixed parameters quantities of the simulation, namely $\tau_f$ and $\ell_f$ prescribing the level of gas turbulence, we can define the dimensionless drift parameter $\Tilde{\epsilon}=\epsilon\tau_f^2/\ell_f$.  
We adopt the fiducial value $\Tilde{\epsilon}=0.1$, to examine the effects of weak but non-negligible turbulence on dust drift. Nonetheless, we extensively explore the influence of this parameter on our results in the discussion, considering values in the range $[10^{-2},10]$. We neglect both particle-particle interactions and their feedback onto the gas.

\section{Results}\label{sec:results}
\begin{figure}[h]
   \centering
   \includegraphics[width=\columnwidth]{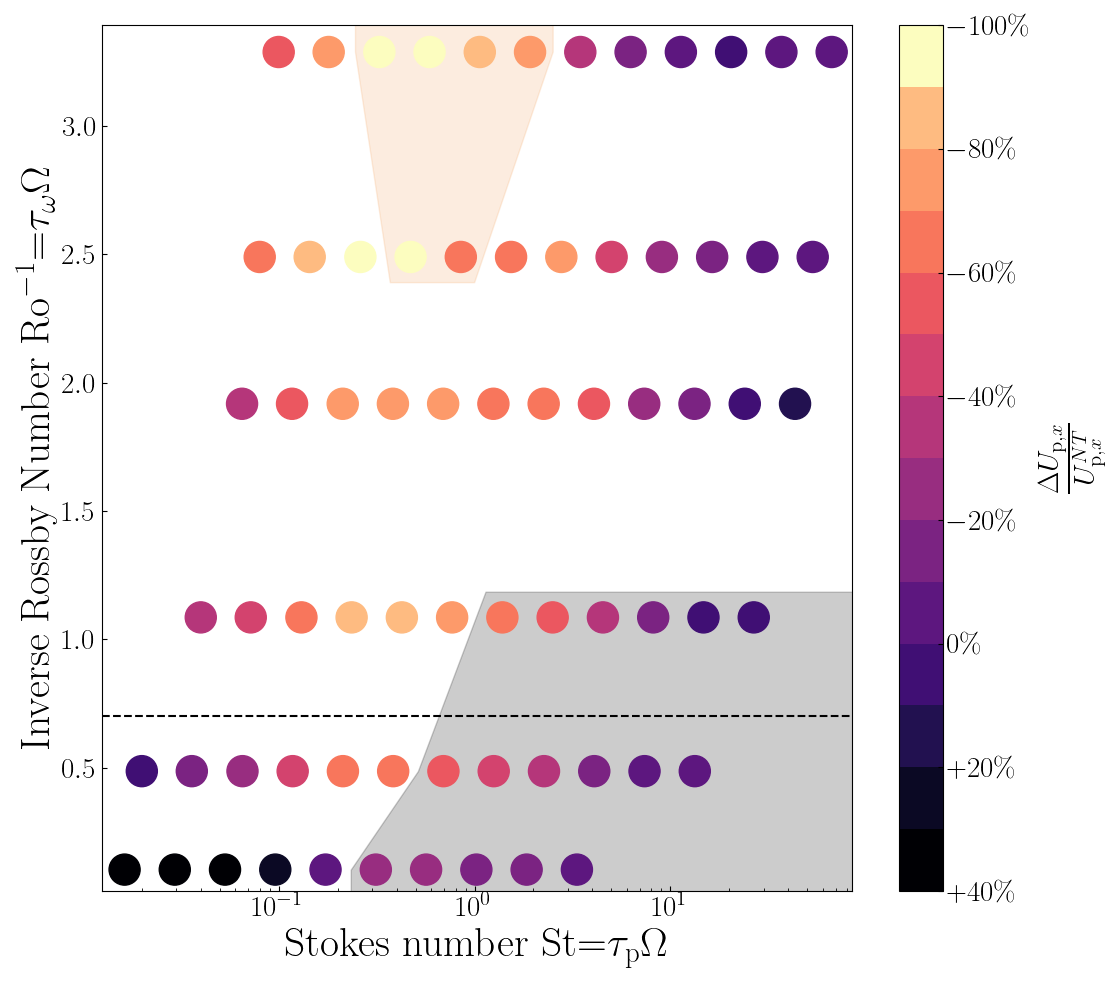}
      \caption{Variation of dust radial drift velocity due to turbulence for $\Tilde{\epsilon}=0.1$ as a function of the inverse Rossby number and the Stokes number.}
         \label{Fig2D}
   \end{figure}

An overview of the results is depicted in Fig. \ref{Fig2D}. The strength of rotation with respect to turbulence is increased along the vertical axis, while the horizontal axis represents the particle Stokes number, varying from small to large grains. 
The orange background zone highlights the region where dust forms point clusters within 
anticyclones, whereas in the gray background area dust fills the whole space. In the other regions of the parameter space, dust particles tend to concentrate
on filaments. For a more detailed discussion of distinct dust concentration regimes, refer to \citet{gerosa2023} or to the highlights given in Appendix \ref{app:cluster}. In the lower region, indicative of strong turbulence and/or slow rotation, dust particles tend to be expelled from eddies. Conversely, in the upper region, dust predominantly resides within anticyclones. The transition between these two behaviors can be quantitatively characterized (see Appendix \ref{app:Ow}) and is identified in Fig. \ref{Fig2D} by the dashed line. To determine whether turbulence diminishes the radial drift of dust, we compare its mean speed in our turbulent runs to that in the laminar case. In the absence of turbulence, the particle radial drift velocity is given by $U_{{\rm p},x}^{\rm NT}=-\tau_{\rm p}\epsilon/(1+\tau_{\rm p}^2\Omega^2)$ (as detailed in Appendix \ref{app:meanv}). We therefore define the relative velocity $\Delta U_{{\rm p},x}/U_{{{\rm p},x}}^{\rm NT}=(\langle U_{{\rm p},x}\rangle-U_{{{\rm p},x}}^{\rm NT})/U_{{{\rm p},x}}^{\rm NT}$, where $\langle\cdot\rangle$ denotes the average over all particles and times. This quantity measures the variation of dust radial drift velocity attributable to turbulence. A negative (positive) value indicates a decrease (increase) in the radial drift of dust compared to the laminar case. It is observed in Fig. \ref{Fig2D} that the radial drift velocity is significantly reduced at large $\Ro^{-1}$ and small Stokes numbers. This indicates that, for these parameters, turbulence substantially slows down dust radial drift. It is noteworthy that, for $\Ro^{-1}=3.29$, the plot is the result of an average over 12 runs.  Because of particles clustering to a point in phase space, many runs are here needed to achieve sufficiently high statistics.

%                               2D radial drift
%-----------------------------------------------------------------
%
%______________________________________________________________

\subsection{Radial drift}
\begin{figure}
   \centering
   \includegraphics[width=\columnwidth]{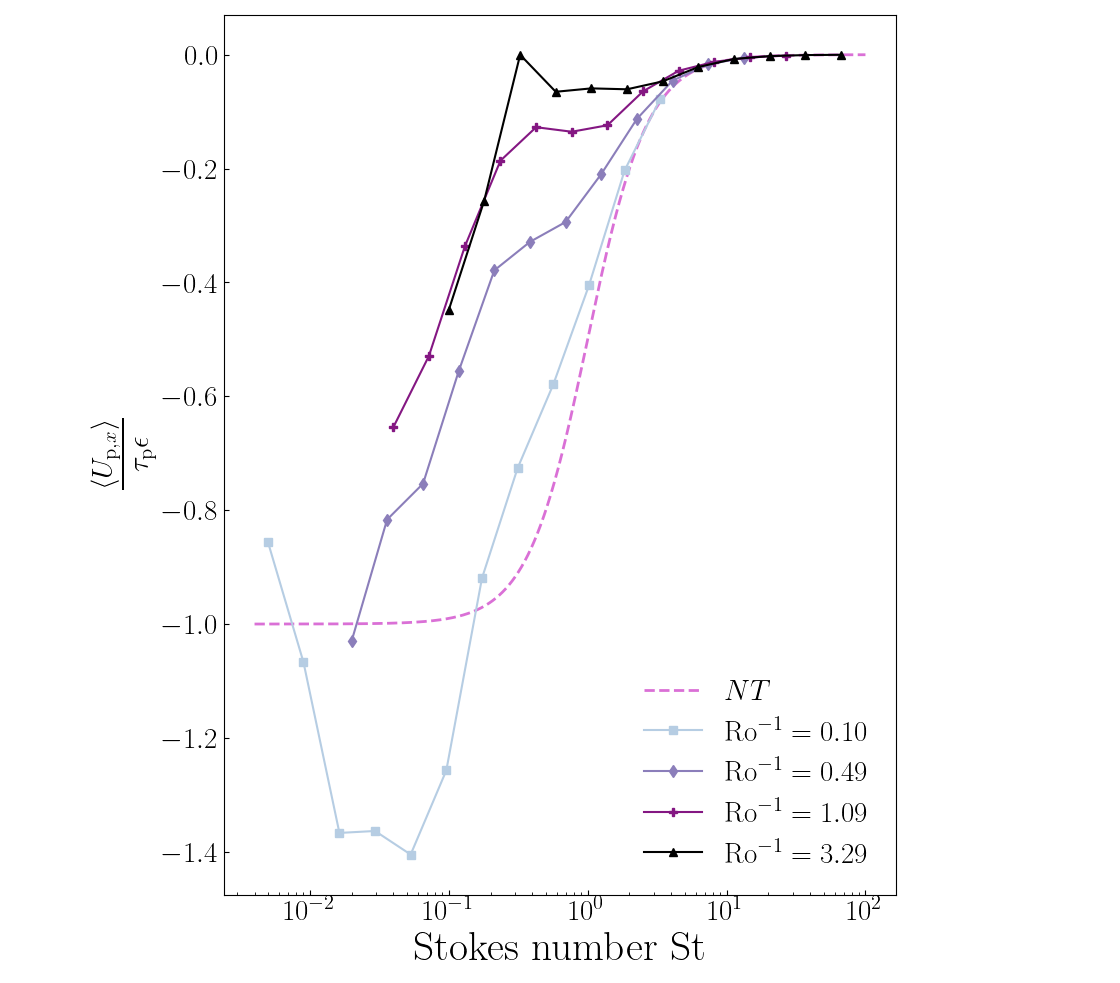}
      \caption{Mean radial drift velocity as a function of the Stokes number, for various values of the Rossby number and for $\Tilde{\epsilon}=0.1$. The dotted line represents the radial drift velocity particles would have in a laminar gas flow.}
         \label{FigRad}
   \end{figure}

The mean radial drift velocity of particles is presented in Fig. \ref{FigRad}. The dashed line corresponds to the non-turbulent case, $U_{{\rm p},x}^{\rm NT}$. For small Stokes numbers, the radial drift velocity exhibits substantial reduction for almost all the Rossby numbers, diminishing with faster rotation rates (larger $\Ro^{-1}$). This can be clearly identified on this figure as the absolute value of the mean radial velocities are smaller than in the laminar case. For $\Ro^{-1}>0.7$, dust lies longer in anticyclones, where it gets locked to the gas radial velocity\footnote{By construction, there is no mean gas radial velocity in the shearing box.}. Turbulent decrease of dust drift reaches up to an order of magnitude in some cases (i.e., circles for which $\Delta U_{{\rm p},x}/U_{{{\rm p},x}}^{\rm NT}=-90\%$ in Fig.~\ref{Fig2D}). Yellow circles in Fig.\ref{Fig2D}, on the other hand, correspond to a complete cessation of radial drift due to turbulence, with particles clustered to a point and trapped inside an anticyclone. For $\Ro^{-1}=0.49$ particles mostly reside between eddies, yet their radial drift is still slowed down by turbulence. For $\Ro^{-1}=0.10$, instead, small particles drift faster toward the star, as usually assumed for strong turbulence. We will delve into the interpretation of these last two phenomena in Sect.~\ref{sec:discussion}. Finally, for all Rossby numbers, $\langle U_{{\rm p},x}\rangle$ converges at large Stokes numbers to $U_{{\rm p},x}^{\rm NT}$, indicating large solids weakly coupled to the turbulent flow.

%                               Radial drift
%-----------------------------------------------------------------
%
%______________________________________________________________

\subsection{Azimuthal drift}
\begin{figure}
   \centering
   \includegraphics[width=\columnwidth]{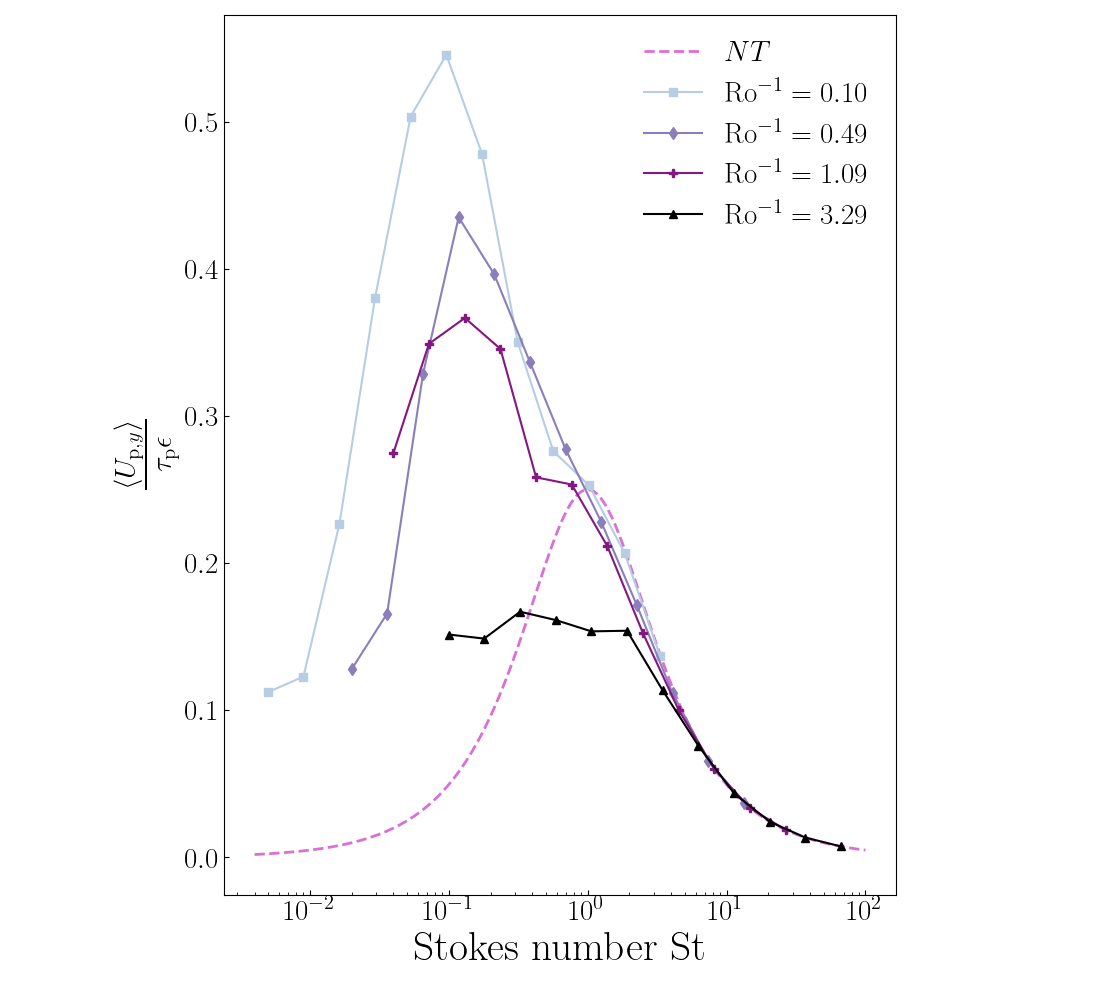}
      \caption{Mean azimuthal drift velocity as a function of the Stokes number, for various values of the Rossby number and for $\Tilde{\epsilon}=0.1$. The dotted line represents the azimuthal drift velocity particles would have in a laminar gas flow.}
         \label{FigAzi}
   \end{figure}

Figure \ref{FigAzi} shows the mean azimuthal drift velocity $\langle U_{{\rm p},y}\rangle$ of dust particles. The laminar drift velocity in the azimuthal direction is given by $U_{{\rm p},y}^{\rm NT}=\tau_{\rm p}^2\Omega\epsilon/2(1+\tau_{\rm p}^2\Omega^2)$ (see Appendix \ref{app:meanv}), plotted as a dashed line. Turbulence enhances the azimuthal drift of small dust particles, while for large $\St$, it remains unaltered. The values of $\langle U_{{\rm p},y}\rangle$ clearly appear larger at slower rotation rates (small $\Ro^{-1}$). It is important to notice that the magnitude of the strongest azimuthal drift velocity represents only a small fraction of the Keplerian velocity ($\langle U_{{\rm p},y}\rangle\approx10^{-4}r_0\Omega$ for $\Ro^{-1}=0.10$ and $\St=0.1$). Therefore, it would not be currently detectable from observational data. Conversely, the azimuthal drift is reduced at very high rotation rates and intermediate Stokes numbers. This is again consistent with the mechanism of dust trapping inside anticyclones, resulting in particle cluster velocities governed by the mean flow.

%                               Azimuthal drift
%-----------------------------------------------------------------
%
%______________________________________________________________

\section{Discussion}\label{sec:discussion}

\subsection{Preferential sweeping}

Preferential sweeping is a well known process that induces variations in particle speed, due to turbulence in the gas phase. This phenomenon has been extensively studied and demonstrated in the context of droplets or aerosols settling in the atmosphere (\citealt{maxey1987}; see also \citealt{bec2024statistical} for a recent review). In such scenarios, the interaction between gravity and turbulence leads to
an acceleration of particle settling, as particles tend to preferentially sample
down-welling regions of the turbulent flow. We anticipate a similar mechanism for dust particles moving within turbulent protoplanetary disks, where gravity drives radial drift of the dust toward the central star. In Fig. \ref{FigRad}, particularly for $\Ro^{-1}=0.10$, one can notice that radial drift is increased by turbulence for intermediate Stokes numbers. A simple analysis using cellular eddies, as depicted in Fig. \ref{Figsnapshots}a, further elucidates this point. Cyclonic eddies (rotating in the direction of the disk) are shown in red, while anticyclonic ones in blue. It is evident that dust particles exclusively sample left-welling regions, attracted by the star's gravity, resulting in increased radial drift velocities. When inspecting snapshots from our numerical simulations (Fig. \ref{Figsnapshots}b and d), the dynamics of dust appears to be more intricate. Here, particles predominantly sample left-welling regions while also undergoing azimuthal drift. During this process, they are once again subject to preferential sweeping. Consequently, they experience an additional acceleration in the azimuthal direction, as seen for small enough Stokes numbers in Fig \ref{FigAzi}.

\subsection{Reduction of radial dust drift}

\begin{figure}
  \centering
  \includegraphics[width=\columnwidth]{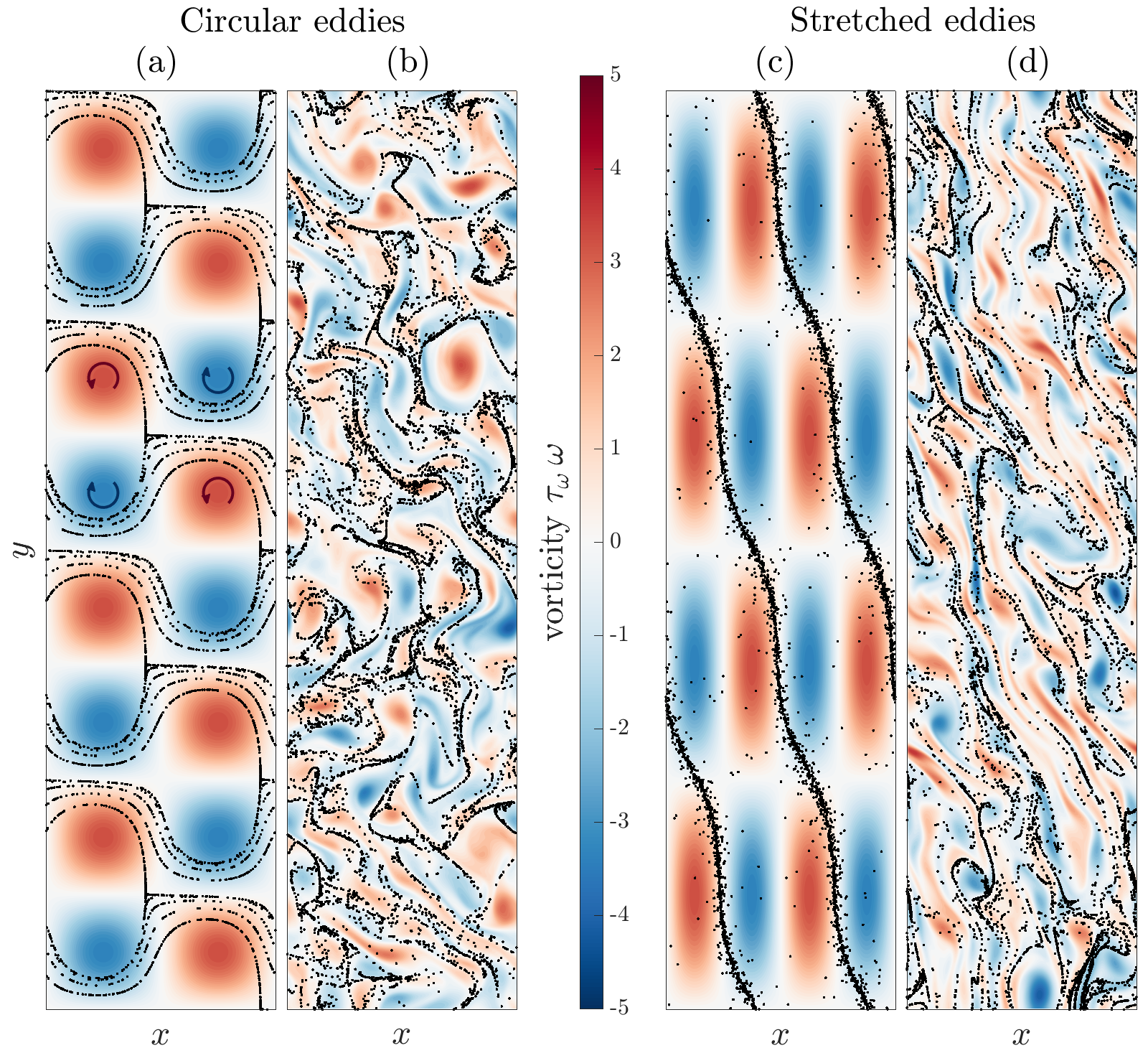}
     \caption{Vorticity $\omega$, normalized by $\tau_{\omega}$, and dust particle position $\vec{R}_{\rm p}$ (black dots) for $\St\approx0.1$. The snapshots in a) and c) have been obtained for a simple cellular flow, while b) and d) from our direct numerical simulations at $\Tilde{\epsilon}=0.1$ and $\Ro^{-1}=0.10$ for b), $\Ro^{-1}=0.49$ for d).
     }
        \label{Figsnapshots}
\end{figure}

Figure \ref{FigRad} shows that, for $\Ro^{-1}>0.10$, the radial velocity of dust particles is decreased by turbulence. Two different regimes should be considered here.
For $\Ro^{-1}>0.7$  particles mostly remain trapped inside anticyclones. Therefore, the reduction of their drift can be readily explained: particles evolve with the radial velocity of eddies, which is slower than the dust velocity. 
On the other hand, at $\Ro^{-1}=0.49$, where particles are instead ejected from eddies, counter-intuitive in regard to the effect of preferential sweeping on dust speed our findings are in contrast with the typical enhancement of dust speed caused by standard preferential sweeping. The reason of this radial drift reduction lies in the Coriolis force, which induces an interchange of azimuthal and radial velocities:
\begin{equation}
   f_{\rm C}= -2\vec{\Omega}\times\vec{U}_{\rm p} = 2\Omega U_{{\rm p},y}\,\vec{e}_x -2\Omega U_{{\rm p},x}\,\vec{e}_y. 
\end{equation}
The decrease of radial drift velocities by turbulence can therefore be 
easily explained. While preferential sweeping would normally enhance the radial velocity of dust, it also increase $U_{{{\rm p},y}}$ that, when entering the Coriolis term, can counteract the acceleration of radial drift. This effect can even prevail, particularly at high rotation rates, resulting in an effective decrease of dust radial drift velocities. Moreover, as rotation rate augments, shear renders
turbulence increasingly non-isotropic \citep{gerosa2023}. In particular, eddies become strongly stretched azimuthally, as can be seen when comparing Fig. \ref{Figsnapshots}b ($\Ro^{-1}=0.10$) and Fig. \ref{Figsnapshots}d ($\Ro^{-1}=0.49$). This, in turn, affects the dynamics of dust. To better understand this mechanism, in Fig.~\ref{Figsnapshots}a and c, we illustrate dust dynamics in a model flow
consisting of cellular eddies with two different aspect ratios to represent vortex stretching due to shear. In both cases, particles follow the preferential sweeping scenario, solely sampling left-welling regions. However, with elongated eddies, dust paths are predominantly azimuthally oriented, thus resulting in a significant reduction in particle mean radial motion. 

%                                       Qualitative approach
%-----------------------------------------------------------------
%
%______________________________________________________________

\subsection{Estimation of the turbulent parameter $\alpha$}
Turbulent intensity is classically defined as the ratio between the amplitude of the turbulent fluctuations $\vec{u}$ of the gas and its mean velocity. This quantity is the most relevant for estimating the effects of turbulence on particle dynamics. From its squared value, we can compute the turbulent $\alpha$ parameter as: 
\begin{equation}
   \alpha=\frac{\langle |\vec{u}|^2\rangle}{(\Omega H)^2}. 
\end{equation}
The modifications of both dust velocities, due to preferential sweeping, and particle concentrations can be estimated through this value. Considering $H=10\,l_{\rm f}$, the $\alpha$ parameter can be as large as $10^{-2}$ for $\Ro^{-1}=0.10$. At this rotation rate, turbulence is ejecting particles from eddies and accelerating their drift. Its effect on particle dynamics is therefore similar to what is already known in the literature for strong, MRI-like turbulence \citep{yang2018}. On the other hand, at $\Ro^{-1}=6.19$ we obtain $\alpha=10^{-5}$. At these low turbulence levels, not extensively studied in the literature, turbulent anticyclones are concentrating dust and slowing down drift. Indeed, in this case, the Coriolis force due to rotation is able to counteract the diffusive power of eddies. We therefore observe a substantial difference in dust dynamics in weak turbulence compared to the purely laminar case  ($\Ro^{-1}=\infty$), where particles would be randomly distributed and would drift radially at a considerably faster velocity.\\
When instead considering the angular momentum transport due to turbulence \citep{shakura1973}, we can compute a ``viscous'' $\alpha$ using the Reynolds stress $\langle u_x\,u_y\rangle$ (in place of $\langle |\vec{u}|^2\rangle$). For this parameter we find values of two orders of magnitude lower than above. This indicates a small amount of turbulent momentum transport in our simulations, in line with recent observations that tend to infer low turbulent viscosity in disks.

\subsection{Dependence on the drift parameter $\Tilde{\epsilon}$}

\begin{figure}[h]
   \centering
   \includegraphics[width=\columnwidth]{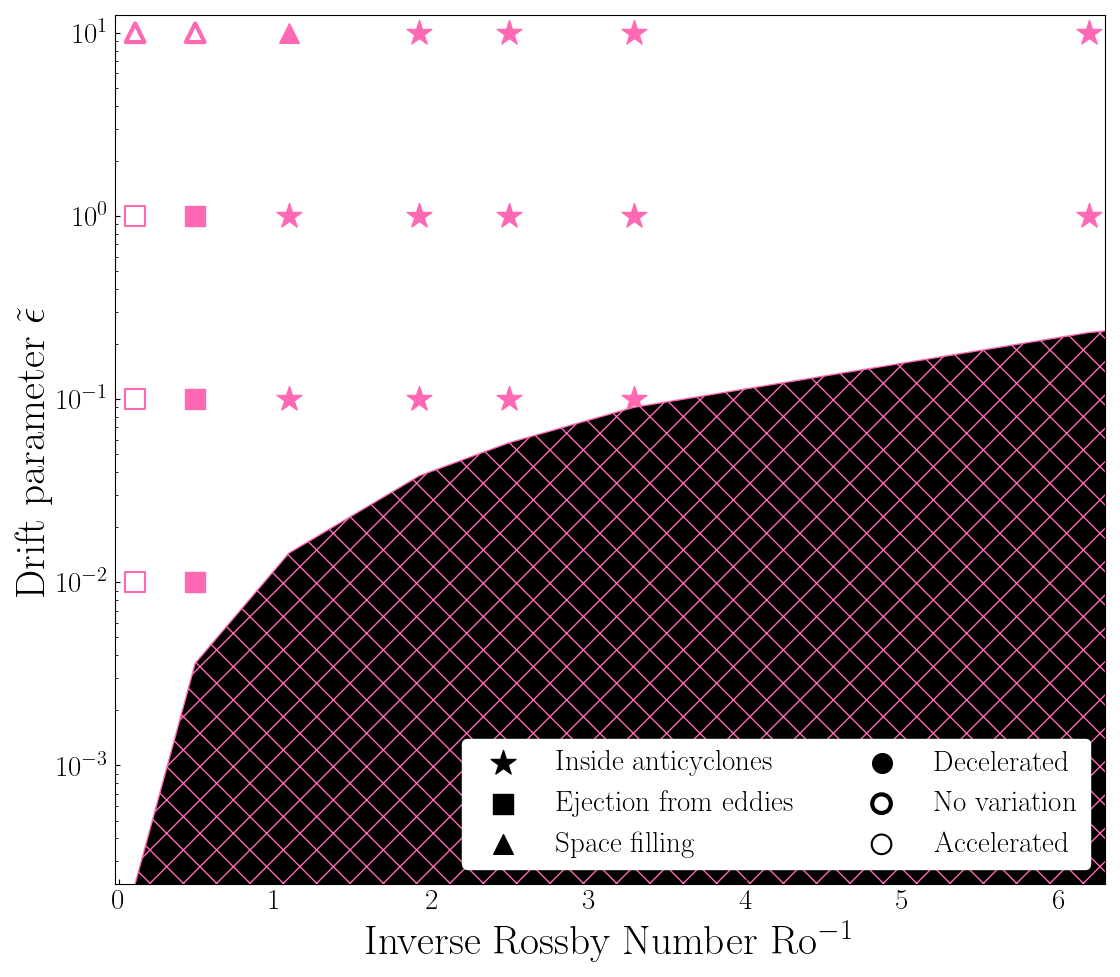}
      \caption{Phase diagram of the drift parameter vs the inverse Rossby number. The highlighted parameters present results on the dust concentration and radial drift for $\St=0.1$.}
         \label{Figepsilon}
   \end{figure}

The shearing box approximation holds true only when the box is located at a distance $r_0$ from the star much larger than the largest scales of the simulation. This requires $r_0\gg l_{\rm f}$. Consequently, we obtain the condition $\epsilon\gg(\Omega_K^2-\Omega^2)\,l_{\rm f}$. The acceptable
values of $(\Omega_K^2-\Omega^2)$ are determined from:
\begin{equation}
{v_y^2}/{v_{\rm K}^2}=1-n\,{c_{\rm s}^2}/{v_{\rm K}^2},
\end{equation}
where $n$ is the index of the power law describing pressure as a function of radius. In the literature, typical disk properties yield $0.996<{v_y^2}/{v_{\rm K}^2}<0.999$. For our analysis, we make the arbitrary choice ${v_y^2}/{v_{\rm K}^2}=0.998$. Therefore, for slow rotation rates, $\Tilde{\epsilon}$ can be as small as $10^{-4}$, while faster rotations necessitate considering larger drift parameter (e.g., $\Tilde{\epsilon}_{\rm min}\approx0.2$ for $\Ro^{-1}=10$). In Fig.~\ref{Figepsilon} the black hatched region shows the inaccessible values of $\Tilde{\epsilon}$ for the shearing box approximation as a function of the inverse Rossby number. Choosing a larger (smaller) value for ${v_y^2}/{v_{\rm K}^2}$ shifts the inaccessible region to smaller (larger) values of the drift parameter $\Tilde{\epsilon}$.

%                                       Parameter space
%-----------------------------------------------------------------
%
%______________________________________________________________
In our exploration of the parameter space, we investigated various values of $\Tilde{\epsilon}$ within the accessible range. Figure \ref{Figepsilon} presents the results on dust concentration and radial velocity variation as a function of $\Tilde{\epsilon}$ and $\Ro^{-1}$ for $\St=0.1$. It demonstrates that, for $\Tilde{\epsilon}\leq 1$, dust is ejected from eddies, and its velocity is enhanced at small $\Ro^{-1}$. Conversely, dust tends to cluster more readily and experiences a decrease in radial drift for $\Ro^{-1}>0.7$. The results at $\Tilde{\epsilon}=1$ further corroborate the findings presented in this paper, as they exhibit the same behavior as those at $\Tilde{\epsilon}=0.1$, while being much further away from the inaccessible region (see Appendix \ref{app:+eps}). However, at $\Tilde{\epsilon}=10$, for which turbulence becomes a secondary mechanism compared to drift, a higher rotation rate is required for the slowdown of dust and particle concentration to happen. It is important to note that the decrease of dust radial drift is quantitatively smaller for the same $\Ro$ and $\St$ at larger $\Tilde{\epsilon}$. Therefore, a small value of the drift parameter is essential for the reduction of radial drift due to turbulence to significantly impact the long-term evolution of dust in disks.

\section{Conclusions}

We addressed in this Letter the issue of dust particle drift in a turbulent gas, usually considered as a problem in theories of planetesimal formation. We demonstrated
that turbulent eddies can reduce the radial drift velocity of small dust particles through two distinct mechanisms: preferential sweeping of particles between elongated eddies at slow
rotation rates and concentration within anticyclones for faster rotations. In the latter scenario,
dust particles may even come to a complete halt in their radial drift.
Concurrently, turbulence tends to enhance the azimuthal drift of dust in most cases, although azimuthal velocities of particles are reduced when they aggregate inside anticyclonic eddies.

The reduction of dust radial drift has significant implications, potentially reconciling the masses of dust disks inferred from observations with the predicted mass loss rates 
from theory, particularly when turbulence prolongs the timescale of dust drift by an order of magnitude. Furthermore, our findings suggest that a turbulent region in the disk could act as a traffic jam, potentially explaining the formation of certain observable substructures, such as those detected by telescopes like ALMA. Slowing down radial drift may also facilitate planetesimal formation by assisting dust particles in overcoming the radial drift barrier and reducing collision speeds, both crucial factors for a favorable outcome of their interactions.

In our simulations, we made several assumptions.
The shearing box approach is relevant for our local numerical simulations, focusing specifically on dust particles dynamics. The turbulence intensity and considered scales are also small enough for density perturbations to be weak in the box \citep{meheut2015}, therefore supporting the incompressible approximation. The choice of two-dimensional simulations is justified for small Rossby numbers, as numerous studies have indicated a two-dimensionalization of the flow at high rotation rates \citep{yeung1998,biferale2016}. Additionally, for large $\Ro$, the 2D approximation remains valid due to the highly stratified nature of protoplanetary disks
\citep{cambon2001}. Finally, we did not consider in this study the back-reaction of dust on 
gas or self-interactions between particles. While these effects can be significant when dust particles are concentrated, they are often considered secondary aspects when  diluted. However, dust feedback can have strong consequences on both particle clustering \citep{johansen2007} and dust radial drift \citep{dipierro2018}, topics that are left for future work.

\begin{acknowledgements}
      This work was supported by the ``Programme National de Physique Stellaire'' (PNPS) and ``Programme national de planétologie'' (PNP) of CNRS/INSU co-funded by CEA and CNES, and Observatoire de la Côte d'Azur. This work was supported by the French government, through the UCA$^\text{JEDI}$ Investments in the Future project managed by the National Research Agency (ANR) under reference number ANR-15-IDEX-01. The authors are grateful to the OPAL infrastructure and the Université Côte d’Azur’s Center for High-Performance Computing for providing resources and support.
\end{acknowledgements}

\bibliographystyle{aa} 
\bibliography{LetterDrift}

\begin{appendix}

\section{Run parameters} \label{app:run}
Table~\ref{tab1} presents the parameters used in the direct numerical simulations with the spectral code Snoopy.
\begin{table}[h!]
    \caption{Simulation parameters (in simulation units).}
    \begin{tabular}{cccccc}
    \hline \hline\\
    $L_x\times L_y$ & $N_x\times N_y$ & $\varepsilon_{\rm I}$ & $\eta_{\rm I}$ & $\nu$ & $\gamma$\\
    \hline\\
    $2\pi\times8\pi$ & $256\times1024$ & $1.38$ & $22.12$ & $0.005$ & $0.05$\\
    \hline\hline
    \end{tabular}
    \label{tab1}
\end{table}

We conducted 7 distinct runs, each varying the rotation rate $\Omega$, resulting in specific values for the eddy turnover time $\tau_\omega$ (as listed in Table~\ref{tab2}), and, consequently, for the Rossby number.
\begin{table}[h!]
    \caption{Variation of the eddy turnover time and Rossby number as a function of rotation (in simulation units).}
    \begin{tabular}{l| l l l l l l l l }
    \hline \hline\\
    $\Omega$ & 0.7 & 2.7 & 5.3 & 8.7 & 10.7 & 13.3 & 21.3 \\
    \hline\\
    $\tau_{\omega}$ & 0.16 & 0.18 & 0.20 & 0.22 & 0.23 & 0.24 & 0.29\\
    \hline\\
    $\Ro^{-1}$ & 0.10 & 0.49 & 1.09 & 1.92 & 2.49 & 3.29 & 6.19\\
    \hline\hline
    \end{tabular}
    \label{tab2}
\end{table}

Finally, each of these runs is seeded with 48 sets of $10^4$ Lagrangian particles each. These families are associated with 4 different values of $\epsilon$ (0.1, 1, 10, 100) and 12 values of $\tau_{\rm p}$ (0.008, 0.014, 0.025, 0.044, 0.080, 0.144, 0.260, 0.470, 0.849, 1.533, 2.769, and 5.000).

\section{Shearing box and drift}\label{app:shearingbox}

We consider dust particles in Keplerian rotation, thus rotating with a rotation rate $\Omega_{\rm K}$ faster than the gas, which rotates with the shearing box at a sub-Keplerian speed $\Omega$.
Considering the box centered at a distance $r_0$ from the star, the equation for passive particles is: 
\begin{equation}
\begin{split}
    \frac{d\vec{V}_{\rm p}}{dt}=&-\frac{1}{\tau_{\rm p}}\left[\vec{V}_{\rm p}-\vec{v}(\vec{R}_{\rm p},t)\right]-2\vec{\Omega}\times\vec{V}_{\rm p}+(2\Omega_K^2+\Omega^2)R_{{\rm p},x}\, 
    \vec{e}_x\\&-(\Omega_K^2-\Omega^2)r_0\,\vec{e}_x.
\end{split}
\end{equation}
For numerical reasons, it is easier to compute the particle velocity in the frame where the background negative shear has been subtracted. Therefore, we perform the change of variables
$U_{{\rm p},y}=V_{{\rm p},y}+(3/2)\Omega x_{\rm p}$ giving:
\begin{equation}
\begin{split}
    \frac{dU_{{\rm p},x}}{dt}=&-\frac{1}{\tau_{\rm p}}\left[U_{{\rm p},x}-u_{x}(\vec{R}_{\rm p},t)\right]+2\Omega U_{{\rm p},y}+(2\Omega_K^2+\Omega^2)\,R_{{\rm p},x}\,\\
    &-(\Omega_K^2-\Omega^2)r_0-3\Omega^2\,R_{{\rm p},x},\,\\ 
     \frac{dU_{{\rm p},y}}{dt}=&-\frac{1}{\tau_{\rm p}}\left[U_{{\rm p},y}-u_y(\vec{R}_{\rm p},t)\right]-\frac{1}{2}\Omega U_{{\rm p},x}.
\end{split}
\end{equation}
Finally, if $R_{{\rm p},x}\ll r_0$, we get $2(\Omega_K^2-\Omega^2)x_{\rm p}\ll(\Omega_K^2-\Omega^2)r_0$, so that
\begin{equation}
    \frac{d\vec{U}_{\rm p}}{dt}=-\frac{1}{\tau_{\rm p}}\left[\vec{U}_{\rm p}-\vec{u}(\vec{R}_{\rm p},t)\right]-2\vec{\Omega}\times\vec{U}_{\rm p}+\frac{3}{2}\Omega U_{{\rm p},x}\,\vec{e}_y-\epsilon\,\vec{e}_x,
    \label{part_eq}
\end{equation}
with $\epsilon=(\Omega_K^2-\Omega^2)r_0$.

\section{Dust clustering}\label{app:cluster}
\begin{figure}[]
   \centering
   \includegraphics[width=\columnwidth]{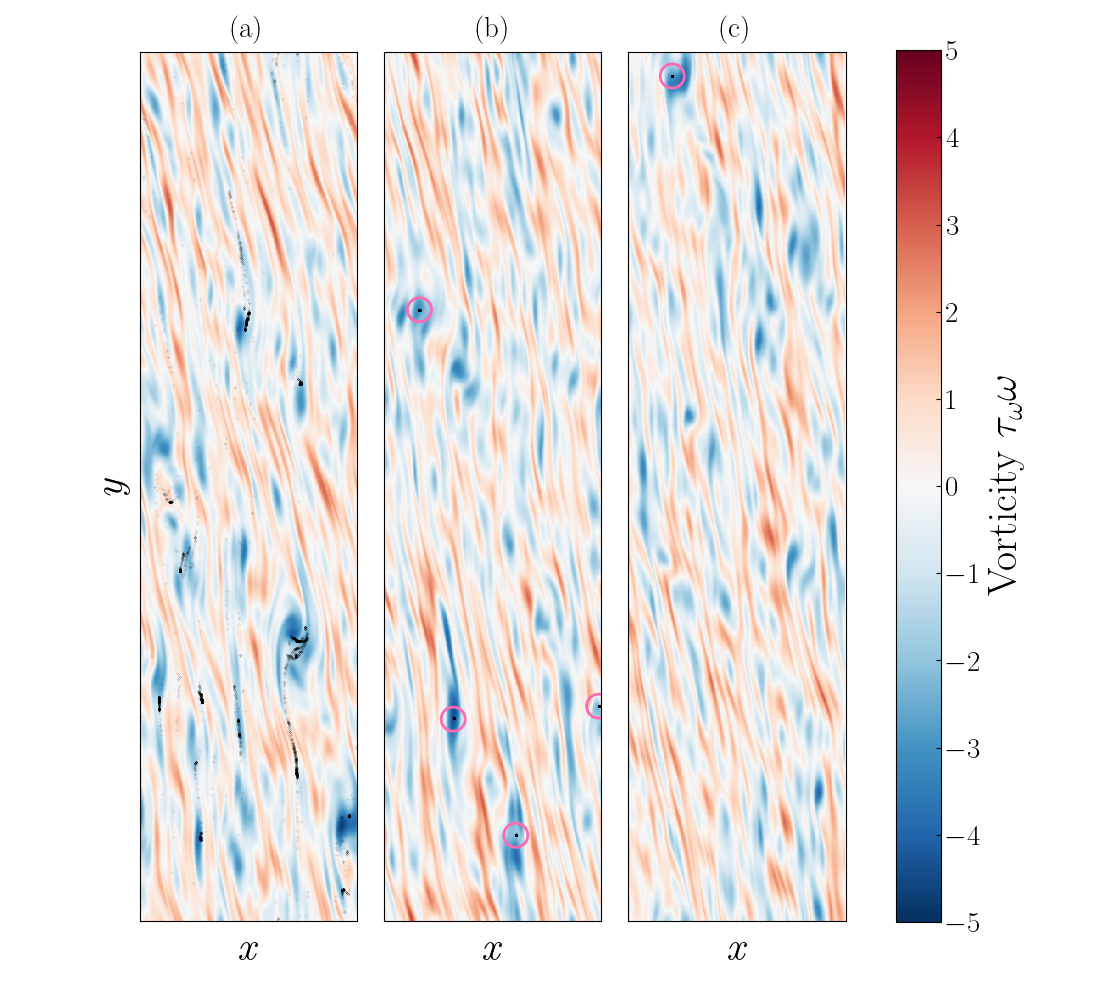}
      \caption{Vorticity $\omega$, normalized by $\tau_{\omega}$, and dust particle position $\vec{R}_{\rm p}$ (black dots) for  $\Tilde{\epsilon}=0.1$,$\St=1$, $\Ro^{-1}=3.29$ and a) t=200$\tau_{\omega}$, b) t=2000$\tau_{\omega}$, c) t=20000$\tau_{\omega}$. The clusters in b) and c) have been highlighted with a surrounding circle.  
              }
         \label{fig:evol}
   \end{figure}

The clustering properties of particles in a turbulent Keplerian flow can vary significantly  depending on the flow and dust parameters. 
In a previous paper where no drift was considered \citep{gerosa2023}, we identified three possible turbulent concentration paths:
\begin{enumerate}
    \item Ejection from eddies and concentration on filamentary structures in between, for small rotation rates and dust sizes.
    \item Distribution in spirals inside the anticyclones, but avoiding their core, for intermediate rotation rates.
    \item Formation of point clusters in the cores of anticyclones, for large rotation rates and small dust sizes.
\end{enumerate}
When the drift is considered, the picture remains qualitatively the same. As an example, we show in Fig. \ref{fig:evol} the concentration of dust particles over time, for $\Tilde{\epsilon}=0.1$, $\St=1$, and $\Ro^{-1}=3.29$. Due to the Coriolis force, dust initially gathers in anticyclonic eddies (Fig. \ref{fig:evol}a). Over time, each of these clouds coalesce into 
a point cluster in the core of an anticyclone (Fig. \ref{fig:evol}b). Clusters 
eventually merge to form a single cluster (Fig. \ref{fig:evol}c).
In the extreme case of $\Tilde{\epsilon}=10$, however, the process of dust clustering is halted by drift, 
which carries particles away from anticyclones on a timescale too fast for dust to concentrate efficiently within them. 

%                                       Formation of clusters
%-----------------------------------------------------------------
%
%______________________________________________________________

\section{Okubo-Weiss parameter}\label{app:Ow}
\begin{figure}
   \centering
   \includegraphics[width=\columnwidth]{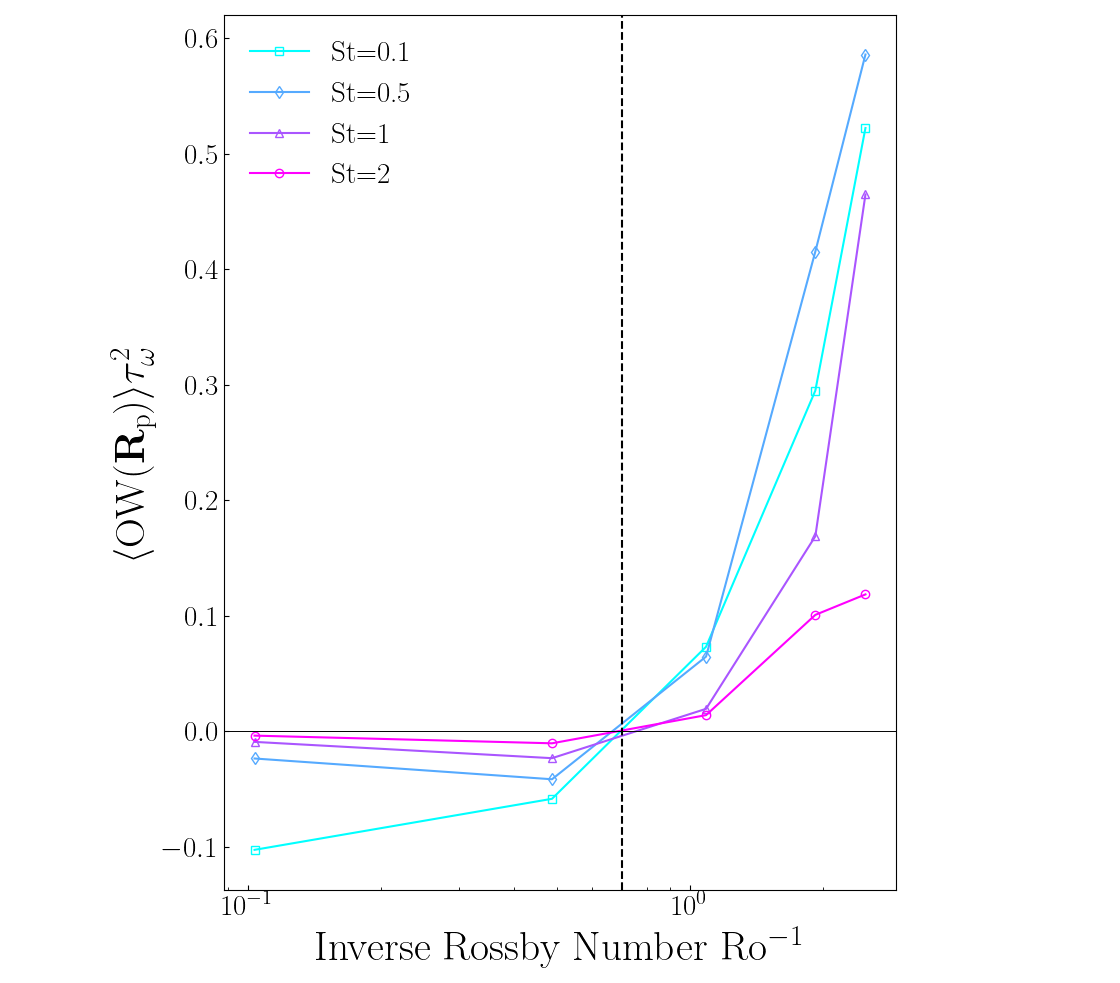}
      \caption{Okubo-Weiss parameter computed at particle position as a function of the inverse Rossby number, for various Stokes numbers and $\Tilde{\epsilon}=0.1$ 
              }
         \label{fig:OW}
   \end{figure}

Okubo-Weiss parameter estimates the properties of a gas flow, helping to identify its local structures. It is defined as:
\begin{equation}
    OW=-(\partial_xu_x)^2-\partial_xu_y\partial_yu_x.
\end{equation}
When positive, it describes rotating, elliptic structures of the flow (e.g., eddies). For straining, hyperbolic regions, the Okubo-Weiss parameter instead assumes negative values. This parameter has been measured along the particles path and its averaged value, as a function of $\Ro^{-1}$ for various Stokes number, is shown in Fig. \ref{fig:OW}. We notice that $\langle OW(\vec{R}_{\rm p})\rangle$ assumes negative values at small $\Ro^{-1}$, while it become positive at high rotation rates. This means that, independently of the Stokes number, particles in a slow rotating gas flow are predominately lying between eddies. Instead, for $\Ro^{-1}\gtrsim0.7$, dust is residing longer inside anticyclonic eddies.

%                                       Okubo Weiss
%-----------------------------------------------------------------
%
%______________________________________________________________

\section{Mean drift velocity}\label{app:meanv}

An expression for the mean particle velocity can be obtained by averaging Eq.~(\ref{part_eq}) assuming statistical stationarity of turbulent fluctuations along particle paths. 
This leads to the following expressions for the mean radial and azimuthal particle velocities
\begin{equation}
\begin{split}
    \langle U_{{\rm p},x}\rangle =& \frac{\langle u_{x}(\vec{R}_{\rm p},t)\rangle+2\tau_{\rm p}\Omega\,\langle u_{y}(\vec{R}_{\rm p},t)\rangle-\tau_{\rm p}\epsilon}{1+\tau_{\rm p}^2\,\Omega^2},\\
    \langle U_{{\rm p},y}\rangle =& -\frac{\tau_{\rm p}\Omega\,\langle u_{x}(\vec{R}_{\rm p},t)\rangle+\langle u_{y}(\vec{R}_{\rm p},t)\rangle-\tau_{\rm p}^2\Omega\,\epsilon}{2(1+\tau_{\rm p}^2\,\Omega^2)}.
\end{split}
\end{equation}
In the absence of gas turbulence ($\vec{u}=0$), these equations yield the mean particle velocity $U_{\rm p}^{\rm NT}$ in a laminar flow.\\
One should note that without drift ($\epsilon=0$), even if the gas flow is turbulent, the average gas velocity at particle position $\langle\vec{u}(\vec{R}_{\rm p},t)\rangle$ would vanish, leading to $\langle \vec{U}_{\rm p}\rangle=0$. On the other hand when drift is present, as in our case, the mean gas velocity at the particle position does not vanish, highlighting the phenomenon of preferential sweeping. This mechanism indeed arises from a preferential sampling of specific zones and structures of the flow along particle paths, modifying the mean radial and azimuthal drifts of the particles.

\section{Radial drift at $\Tilde{\epsilon}=1$}\label{app:+eps}

\begin{figure}
   \centering
   \includegraphics[width=\columnwidth]{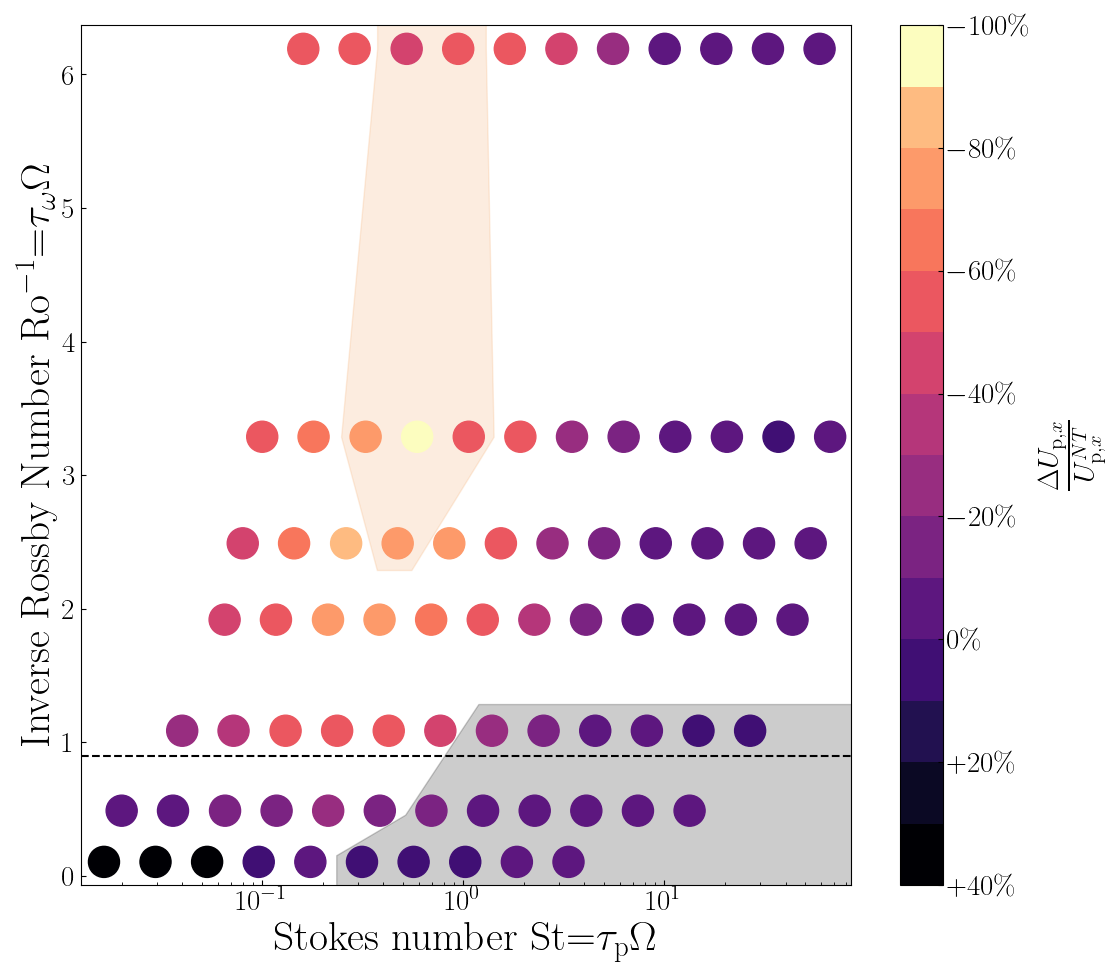}
      \caption{Variation of dust radial drift velocity due to turbulence for $\Tilde{\epsilon}=1$ as a function of the inverse Rossby number and the Stokes number.}
         \label{fig:eps1}
   \end{figure}

Figure \ref{fig:eps1} shows the relative increase of dust radial drift due to turbulence $\Delta U_{{\rm p},x}/U_{{{\rm p},x}}^{\rm NT}=(\langle U_{{\rm p},x}\rangle-U_{{{\rm p},x}}^{\rm NT})/U_{{{\rm p},x}}^{\rm NT}$ for $\Tilde{\epsilon}=1$. We notice that the results closely resemble those presented in Fig.~\ref{Fig2D} for $\Tilde{\epsilon}=0.1$. A subtle reduction of radial drift can be noted at $\Ro^{-1}=6.19$ and small $\St$ compared to $\Ro^{-1}=3.29$, indicating a reversal in trend. In this case, turbulence is too weak to efficiently counteract dust drift. 
\\

%                        2D radial drift
%-----------------------------------------------------------------
%
%______________________________________________________________

\end{appendix}

\end{document}